\documentclass[prd,aps,
showpacs,
twocolumn,
nofootinbib,
floatfix,
superscriptaddress]{revtex4}
\usepackage{url}
\usepackage{hyperref}
\usepackage{graphicx}
\usepackage{epsfig}
\usepackage{rotating}
\usepackage{amssymb}
\usepackage{dsfont}
\usepackage{psfrag}
\usepackage{amsmath,euscript,array,mathrsfs}
\usepackage{mathtools}

\newcommand{\beq}{\begin{equation}}
\newcommand{\eeq}{\end{equation}}
\newcommand{\beqs}{\begin{eqnarray}}
\newcommand{\eeqs}{\end{eqnarray}}

\begin{document}
\title{Constrained Flavor Breaking}

\author{Thomas Appelquist}

\affiliation{Department of Physics, Sloane Laboratory, Yale University, New Haven, Connecticut 06520, USA}

\author{Yang Bai}

\affiliation{Department of Physics, University of Wisconsin, Madison, WI 53706, USA}

\author{ Maurizio Piai}

\affiliation{Swansea University, College of Science,
Singleton Park, Swansea, Wales, UK}


\begin{abstract}
We explore predictive flavor models based on subgroups of the standard-model $SU(3)^5$ flavor symmetry. Restricting to products of $SU(3)$, we find that a global $SU(3)^3$ flavor symmetry, broken only by two Yukawa spurions, leads to a relation among down-type quark, up-type quark and charged-lepton Yukawa matrices: $Y_d \propto Y_u Y_e^\dagger$. As a result, the charged-lepton mass ratios are expressed in terms of quark mass ratios and mixing angles. Large leptonic mixing angles appear to be natural and lead to contributions to flavor-changing neutral currents in the charged-lepton sector, which can be tested in future precision experiments.
\end{abstract}


\maketitle


\section{Introduction}
The standard model (SM) of particle physics  provides a very successful description of physics up to the TeV scale. In particular, the simplest realization of electroweak symmetry breaking, in which the Higgs mechanism is implemented via one or more scalar fields, is in excellent agreement with the properties of the recently discovered Higgs particle~\cite{ATLAS,CMS}. The masses of all the elementary fermions arise after electroweak symmetry breaking due to the presence of Yukawa interactions. The fundamental origin of the Yukawa couplings remains unknown. All attempts at deriving them from a more fundamental theory of flavor involve the observation that in the absence of the Yukawa couplings, the SM Lagrangian possesses an enlarged $G = SU(3)^5$  global symmetry.~\footnote{The symmetry  can be extended to $G=SU(3)^6$ if one assumes the existence of 3 right-handed neutrinos. We do not do so here. There are also a set of $U(1)$ symmetries to consider. Among these, we make direct use of only lepton number and baryon number conservation.} This symmetry is ultimately responsible for the naturalness of the Yukawa couplings and fermion masses, and  it is the origin of the suppression of flavor-changing neutral current (FCNC) transitions.

There exists a vast literature on models of flavor generation, differing from one another by what subgroup of $G$ plays a role in the underlying theory of flavor~\cite{SU3,Feruglio:2015jfa}, and what mechanism(s) yield its breaking. At one extreme sit models with minimal flavor violation (MFV)~\cite{Chivukula:1987py}, in which one assumes that in the ultraviolet (UV) completion of the SM, the $SU(3)^5$ symmetry is violated only by the SM Yukawa couplings. Hence all FCNC processes are suppressed by combinations of the Yukawa couplings (see for example~\cite{MFV}). An advantage of MFV is that it yields a restrictive structure for the possible operators controlling FCNC transitions. As a result the mass scale associated with these processes can be relatively low, in the TeV range.

The MFV framework does not, however, yield any new testable predictions for the mass and mixing matrices appearing in the SM. Hence, it does not reveal anything new about the origin of flavor. The masses, mixing angles, and CP-violating phases are treated as independent free parameters.

In this paper, we investigate whether it is possible to gain some insight into the origin of flavor in an MFV-like framework by constraining more tightly the Yukawa-coupling matrices that break the flavor symmetry.
Our constrained-flavor-breaking (CFB) approach 
is presented as 
an effective field theory 
(EFT) valid only up to the flavor-breaking scale $F$. 
We obtain two relations for charged-lepton mass ratios in terms of the quark masses and mixing angles. 
Using  the smallness of both CKM mixing angles and quark mass ratios,
the symmetry structure enforces large off-diagonal entries 
in the lepton mass matrix,
which suggest the natural appearance of large leptonic mixing angles.

We describe our framework in section~\ref{sec:framework}, and our specific model in section~\ref{sec:model}, focusing on the terms in the EFT Lagrangian responsible for breaking the underlying flavor symmetry and generating the quark and charged lepton masses. We then explore the resultant relations among the Yukawa matrices. We discuss higher-dimension operators responsible for FCNC transitions in section~\ref{sec:fcnc}, coming to the conclusion that leptonic FCNC processes constrain the scale of the higher-dimension operators to be above $10^4$ TeV. We summarize our results and discuss open questions in section~\ref{sec:conclusions}. In appendix~\ref{sec:notation}, we present some notation and list the measured values of quark and lepton masses and CKM mixing angles.

\section{Framework}
\label{sec:framework}

We consider only the quark and charged-lepton sectors, leaving neutrino mass and mixing for future study. The maximal global non-Abelian flavor symmetry in the absence of mass is then $SU(3)^5$. We break the flavor symmetries with dimensionless spurion fields (Yukawa matrices) that are bi-fundamentals under two of the $SU(3)$'s. A minimum of three spurions would be required to break all five of the $SU(3)$'s: two ($Y_d$ and $Y_u$) to break $SU(3)_{Q_L}\times SU(3)_{d_R}\times SU(3)_{u_R}$ and one ($Y_e$) to break $SU(3)_{L_L} \times SU(3)_{e_R}$. As a result, there are no predictions for masses and mixing angles with $SU(3)^5$.

A predictive framework can allow at most two bi-fundamental spurions. This suffices to give mass to all the fermions if there are at most four global $SU(3)$'s, with at least two of the five fermion fields transforming under the same $SU(3)$. With four $SU(3)$'s one spurion can break $SU(3)_{L_L} \times SU(3)_{e_R}$, while the other breaks $SU(3)_{Q_L}\times SU(3)_{u_R, d_R}$. This model, though, predicts $Y_d \propto Y_u$, giving incorrect quark mass ratios and vanishing CKM mixing angles. Other assignments for  $SU(3)^4$ can also be seen 
to yield  unrealistic results.

We therefore turn to a global $SU(3)^3$ symmetry, with two spurions for the flavor breaking. To have realistic quark masses and mixing angles, we assign $Q_L$, $d_R$ and $u_R$ each to a different one of the $SU(3)$'s. There are then six options to assign $L_L$ and $e_R$ to two different $SU(3)$'s.  

For the four cases with either $L_L$ or $e_R$ transforming under $SU(3)_{Q_L}$, one has the unrealistic prediction that the charged-lepton mass ratios are proportional to either the up-type or down-up quark mass ratios. The remaining two cases correspond to $SU(3)_{Q_L}\times SU(3)_{d_R, L_L} \times SU(3)_{u_R, e_R}$ or $SU(3)_{Q_L}\times SU(3)_{d_R, e_R}\times SU(3)_{u_R, L_L}$. 

\begin{figure}[th!]
\begin{center}
\includegraphics[width=0.35\textwidth]{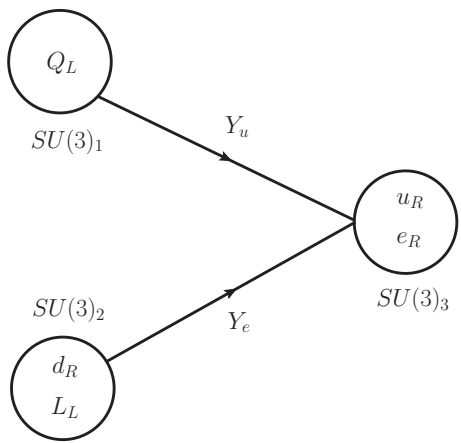}
\caption{A moose diagram for our model.}
\label{Fig:moose}
\end{center}
\end{figure}

In this paper, disregarding the neutrinos and focusing on only the charged-lepton mass ratios in the leptonic sector, these two cases are equivalent because the charged-lepton mass matrices are the Hermitian transpose of each other. We adopt the assignment $SU(3)_{Q_L}\times SU(3)_{d_R, L_L} \times SU(3)_{u_R, e_R}$, as shown in the moose diagram of Fig.~\ref{Fig:moose} and  in Table~\ref{table:fields}, where the three $SU(3)$'s are labeled  $1$, $2$, and $3$.

We next choose the $SU(3)$  assignments for the two bi-fundamental spurions, which we take to be SM singlets.
The  possibility we  focus on, shown in Table~\ref{table:fields}, is to employ a $Y_u$ transforming under $SU(3)_{1} \times SU(3)_{3}$ and a
$Y_e$ transforming under $SU(3)_{2} \times SU(3)_{3}$. This provides independent mass matrices for the up-type quarks and charged leptons. The down-type quark mass matrix will require an EFT term with a product such as $Y_{u}Y_{e}^{\dagger}$, leading to a relation among the lepton and up-type quark mass ratios and CKM mixing angles.

There are two other choices for the two bi-fundamental spurions.  One, with  $Y_u$ transforming under $SU(3)_{1} \times SU(3)_{3}$ and  $Y_d$ transforming under $SU(3)_{1} \times SU(3)_{2}$, would lead to the charged lepton mass matrix arising from a product such as $Y_{d}^\dagger Y_{u}$. In light of the small mixing angles in the CKM matrix, this gives the unrealistic result of the charged-lepton masses being more hierarchical than the up-type quark masses. A third choice, with  $Y_d$ transforming under $SU(3)_{1} \times SU(3)_{2}$ and  $Y_e$ transforming under $SU(3)_{2} \times SU(3)_{3}$, also has a difficulty of obtaining a large enough top Yukawa coupling, at least in models with a single Higgs doublet.~\footnote{All these relations are for ($d_R$, $L_L$) and ($u_R$, $e_R$) having the same representation. For the model with ($d_R$, $L_L$) and ($u_R$, $e_R$) having the complex-conjugate representations, one needs to replace $Y_e$ by $Y_e^*$. Since this  changes only the phases in the lepton sector, predictions of charged-lepton mass ratios are unaffected.}

A restriction to fewer than three global flavor $SU(3)$'s leads to an over constrained flavor structure. In Ref. \cite{GCIW}, for example, two $SU(3)$'s are employed for the quarks and charged leptons, allowing for compatibility with $SU(5)$ grand unification:  one for the $\bar{5}$ ($L_L$, $d_R^c$) and one for the $10$ ($Q_L$, $u_R^c$, $e_R^c$). This leads to quark and lepton mass relations far from the experimental values, addressed in Ref. \cite{GCIW} by the introduction of additional spurions. Our approach, with three flavor $SU(3)$'s provides a more realistic description of fermion masses and mixing angles, but it isn't yet clear to us how to embed it into a larger framework of gauge unification.

The origin of the matrices $Y_u$ and $Y_e$ is not specified here. One could imagine, for example, that $Y_u$ emerges from an underlying dimension-five operator of the form $\eta_u\widetilde{H}\bar{Q}_L\Phi_{u}u_R / F$, where $\widetilde{H}$ is the conjugate of the SM Higgs doublet,
$\eta_u$ is a dimensionless coupling and $\Phi_{u}$ is a dimension-one dynamical field with a vacuum expectation value (VEV) of order the flavor breaking scale $F$. A similar term $\eta_e{H}\bar{L}_L\Phi_{e}e_R / F$ could lead to the charged-lepton mass matrix. The down-quark mass matrix could emerge from an even higher dimension operator, the simplest example being $\eta_d H\bar{Q}_L \Phi_{u} \Phi_{e}^{\dagger}d_R /F^2$.

\section{The Model and Fermion Mass}
\label{sec:model}

The ingredients of our model with their assignments under the global
flavor $SU(3)^3$ are shown in Table~\ref{table:fields}. The list includes the quark and charged-lepton fields and the two dimensionless, bi-fundamental spurion fields $Y_u$ and $Y_e$.
The SM gauge fields and symmetries are not explicitly shown. The focus of this paper is on
flavor breaking, not electroweak breaking, but for simplicity, for most of our
discussion, we take electroweak breaking to be described by a single Higgs
doublet $H\sim(1,2)_{1/2}$ under the SM $SU(3)_{\rm QCD} \times SU(2)_W \times U(1)_Y$ gauge symmetry. 
In this case, $H$ develops the VEV
\beqs
\langle H\rangle &=&\frac{v_W}{\sqrt{2}}\left(\begin{array}{c} 0 \cr 1
\end{array}\right)\,,
\eeqs
where $v_W\simeq 246$ GeV.
We could equally well employ two Higgs doublets which,
for large $\tan{\beta}$ (the ratio between the two vacuum expectation values),
leads to a bottom-quark coupling comparable to the top-quark coupling. In either approach, the important question of the stability of the electroweak scale arises. We will not discuss it further in this paper as our focus is on the physics of flavor-symmetry breaking.

\begin{table}
 \renewcommand{\arraystretch}{1.2}
  \centering
\begin{tabular}{|c|c|c|c|}
\hline\hline
 & $SU(3)_1$ & $SU(3)_2$ & $SU(3)_3$  \cr
\hline
$H$ & 1 & 1 & 1 \cr
\hline
$Q_L$ & 3 & 1 & 1 \cr
$d_R$ & 1 & 3 & 1 \cr
$u_R$ & 1 & 1 & 3 \cr
$L_L$ & 1 & 3 & 1 \cr
$e_R$ & 1 & 1 & 3 \cr
\hline
$Y_u$ & 3 & $1$ & $\bar{3}$ \cr
$Y_e$ & 1 & 3 & $\bar{3}$ \cr
\hline\hline
\end{tabular}
\caption{The global flavor symmetries and field content of the model.}
\label{table:fields}
\end{table}

The quark and charged-lepton mass terms of the EFT are given by
\beqs
\label{Eq:yukawa}
\,-\,H\,\bar{Q}_L  Y_d d_R
\,-\,\widetilde{H}\,\bar{Q}_L Y_u u_R
\,-\,H\,\bar{L}_L  Y_e e_R
\,+\,\cdots\,,
\eeqs
where the dots represent higher-dimensional operators, and where the matrix $Y_d$ must be constructed from our two spurion fields $Y_u$ and $Y_e$. The simplest structure based on the flavor symmetry is
\beqs
\label{Eq:relation}
Y_d &=& \eta\,Y_u Y_e^{\dagger}\,,
\eeqs
where $\eta$ is a parameter to be chosen so that the largest entry in $Y_d$ is $O(10^{-2})$ in order to fit the $b$-quark mass.
This relation may appear counter-intuitive given that the mass hierarchy among the up-type
quarks is much larger than the down-type quarks

An immediate question is whether this relation is  stable in the face of quantum corrections. 
One-loop and higher corrections in the SM, for example from emission and re-absorption of the Higgs boson, 
will correct it, generating additive down-quark mass terms with matrix 
structures containing additional factors such as $Y_{u} Y_{u}^{\dagger}/(4\pi)^2$. 
These terms violate the matrix relation Eq.~(\ref{Eq:relation}) and they typically enter 
with logarithmic UV cutoff dependence. 
But since no element of $Y_{u}$ is larger than unity, these 
are small corrections for any reasonable cutoff. We return to this 
discussion in Section IV, after introducing the higher-dimension 
operators of our EFT, which  correct the dimension-$4$ terms of the SM 
Lagrangian.

After electroweak symmetry breaking, the mass matrices of the quarks and charged leptons are given by
\beq
M^{(d,u,e)}\,=\,\frac{1}{\sqrt{2}}\, Y_{d,u,e}\,v_W\,.
\eeq
We can make $SU(3)_1$ and $SU(3)_2$ transformations to
 the basis in which $Y_d$ is diagonal. Making further use of $SU(3)_3$ and using
 the convention in Eq.~(\ref{Eq:CKM}), there is a basis in which  we can write
\beqs
Y_d&=&{\rm diag}\,(Y_d)\,,~~~~~~~~~
Y_u\,=\,V_{\rm CKM}^{\dagger}\,{\rm diag}\,(Y_u)\,.
\eeqs
Using Eq.~(\ref{Eq:relation}) we then have
\beqs
Y_e&=&\frac{1}{\eta}\,{\rm diag}(Y_d) \,V_{\rm CKM}^\dagger\,  \left[{\rm diag}(Y_u)\right]^{-1}\,.
\label{Eq:mainformula}
\eeqs
Using the central  values of the CKM angles, phase and quark mass parameters
listed in Appendix~\ref{sec:notation} (and defined at the common scale $M_Z$), we find that in this special basis
\beqs
|Y_e| &\simeq&\frac{1}{\eta}\left(\begin{array}{ccc}
1.8 & 0.0011 & 1.4 \times 10^{-7}\cr
8.2 & 0.093 & 0.000014 \cr
6.1 & 0.19 & 0.017
\end{array}\right)\,.
\label{Eq:ye-abs}
\eeqs
The 21 entry of $|Y_e|$, given  by $(1 / \eta) |V_{\rm CKM}^{12}| m_s / m_u$, 
is the largest. We thus infer that
\beqs
m_{\tau}\simeq\,|Y_e|_{21}\,\frac{v_W}{\sqrt{2}}&\simeq&
\frac{1}{\eta}\times (1.4\times 10^3\,{\rm GeV})\,,
\eeqs
requiring $ \eta \simeq 10^{3}$ to fit the experimental $\tau$ mass.
~\footnote{The large value of $\eta$ must emerge from a UV completion of
this model. Suppose, within the framework mentioned at the end of
Section~\ref{sec:framework}, that $Y_e=\eta_e \langle \Phi_e \rangle/F$, with the largest
entry of $\langle \Phi_e \rangle$ of $O(F)$ and $\eta_e \simeq 10^{-2}$.
The matrix $Y_d$ would similarly come from the operator $\eta_d\,H\bar{Q}_L
\Phi_{u} \Phi_{e}^{\dagger}d_R /F^2$. Then, with the assumption of vacuum
saturation, $Y_d = \eta_{d} \langle \Phi_u\rangle  \langle \Phi_e^{\dagger} \rangle$.
Here, the largest entry of $\langle \Phi_u \rangle$ is of $O(F)$, but the largest entry of
the matrix product $\langle \Phi_u \rangle \langle \Phi_e^{\dagger} \rangle$ is of
$O(10^{-3} F^2)$. Then $\eta_d \simeq \eta\, \eta_e \simeq O(10)$.
\\ 
It is also the case that corrections to Eq.~(\ref{Eq:relation}) will 
emerge from whatever theory UV completes our EFT. Suppose, for example, 
that the $\Phi_u$ and $\Phi_e$ fields represent dynamical degrees of 
freedom within the UV completion. A naive estimate would lead to a 
correction of the form $Y_d \propto [Y_u + \epsilon_n (Y_u 
Y_u^\dagger)^n Y_u ] Y_e^\dagger$, with $\epsilon_n = 
O\left\{\left[\frac{\eta_d M}{\eta_u(4\pi)^2 F }\right]^{2n}\right\}$, where $M$ is 
the natural mass scale (cutoff) of the UV completion. 
 This would substantially modify only the $33$ element of $Y_u$ entering 
Eq.~(\ref{Eq:relation}) because of 
the hierarchical structure of $Y_u$.
As long as $M < 4 \pi F$, even the $33$ element
will receive only a small correction.\label{footnote:eta}}

Despite the smallness of the CKM mixing angles, the $Y_e$ matrix has some very large off-diagonal entries due to the large ratios $m_s/m_u$ and $m_b/m_u$. The presence of $[{\rm diag}(Y_u)]^{-1}$ in Eq.~(\ref{Eq:mainformula}) magnifies
the CKM angles. This structure provides an unusual source for large mixing angles in a flavor model, and
 suggests that the large measured values of the PMNS angles are natural. 
 We haven't discussed the neutrino mass matrix  here, 
 yet it would require a fine-tuned relation between $Y_e$ and the neutrino mass matrix to avoid the presence of large mixing angles.

Most importantly, Eq.~(\ref{Eq:mainformula}) leads to relations for the ratios of the charged-lepton masses. Diagonalizing the matrix $Y_e$ and again using the central values of the CKM angles and quark masses, we obtain
\beqs
\label{Eq:theoryrelation1}
\frac{m_e}{m_{\tau}}({\rm th})&\simeq&2.1\times 10^{-4}\,,\\
\frac{m_{\mu}}{m_{\tau}}({\rm th})&\simeq&0.013\,,
\label{Eq:theoryrelation2}
\eeqs
to be compared with the experimental values
\beqs
\frac{m_e}{m_{\tau}}({\rm exp})&\simeq&2.8 \times 10^{-4}\,,\\
\frac{m_{\mu}}{m_{\tau}}({\rm exp})&\simeq&0.059\,,
\eeqs
computed using the renormalized masses defined at the common scale $M_Z$ (see Appendix~\ref{sec:notation}).

\begin{figure}[ht]
\begin{center}
\begin{picture}(250,170)
\put(10,15){\includegraphics[height=4.6cm]{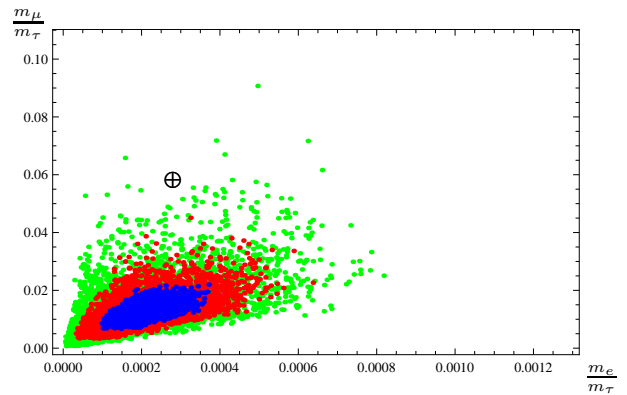}}
\put(5,146){$\frac{m_{\mu}}{m_{\tau}}$}
\put(222,12){$\frac{m_{e}}{m_{\tau}}$}
\end{picture}
\caption{Charged-lepton mass ratios in our model compared to the experimental data (black).
The points are obtained by randomly choosing 4000 possible values
of the masses of the  quarks and CKM angles
 within the $3\sigma$ (green),
$2\sigma$ (red) and $1\sigma$ (blue) ranges. Note that this scatter plot should not be interpreted as the result of a global fit.}
\label{Fig:plotratios}
\end{center}
\end{figure}

The experimental values are quite precise, but the theoretical relations are affected by the comparatively large errors in the measured values of the quark masses, particularly the light ones, the role of which is quite important.
The central-value relation for $m_e/m_{\tau}$ is well within these uncertainties, while the central-value prediction of the muon mass appears to be within a factor of three of the experimental value.
To be more precise, we have numerically varied the masses of the quarks and the
CKM angles within the allowed experimental ranges, and compared the theoretical relations for the charged lepton mass ratios with the experimental data. This is illustrated in Fig.~\ref{Fig:plotratios}.

 To provide some insight into our theoretical relations Eqs.~(\ref{Eq:theoryrelation1},\ref{Eq:theoryrelation2}), in particular their dependence on the quark mass ratios, we use the Wolfenstein parametrization of the CKM matrix Eq.~(\ref{Eq:Wolfenstein}). Expressing the parameters $\rho$ and $\eta$ as $O(1)$ numbers times $\lambda \approx 0.23$ (approximately the Cabibbo angle),
 and taking the quark masses to be within roughly a factor of 2 of their experimental values,  
 we find the approximate formulas

\beqs
\frac{m_e}{m_{\tau}} &\simeq&  \frac{m_d}{m_s} \, \frac{m_u}{m_t}\,\lambda^{-4}  \,,
\label{Eq:approx-formula-1}
\\
\frac{m_\mu}{m_{\tau}}  &\simeq& \frac{m_b}{m_s} \, \frac{m_u}{m_c} \, \lambda \,.
\label{Eq:approx-formula-2}
\eeqs
We can see that to bring $m_\mu/m_\tau$ closer to the experimental value would require increasing $m_u$ 
or decreasing $m_c$ while keeping the product $m_d m_u$ intact.

We are encouraged by these results. Large leptonic mixing angles appear to be natural. The charged-lepton mass ratios emerge from Eq.~(\ref{Eq:mainformula}), in which CKM mixing plays an intricate and prominent role. In the absence of this mixing, the ratio $m_{\mu}/m_{\tau}$ would be in good agreement with experiment, but $m_e/m_{\tau}$ would be predicted to be too large by more than an order of magnitude.

\section{Higher Dimension SM Operators and Flavor-Changing Neutral Currents.}
\label{sec:fcnc}

In the standard model, FCNC transitions are forbidden at the tree level due to the GIM mechanism,
and arise only at the one-loop level. Due to the one-loop, weak-coupling suppression, the smallness of the CKM off-diagonal terms, and the quark mass hierarchies, all such processes are strongly suppressed. The underlying reason for the suppression is $SU(3)^5$ flavor symmetry,  the breaking of which requires diagrams with the insertion of both the up-type and down-type Yukawa couplings. While this is an automatic consequence of the SM assignments of the fermions,
the GIM mechanism is not a generic feature of extensions of the standard model.

This potential difficulty has been addressed by the MFV framework \cite{MFV} in which \emph{all} breaking of the $SU(3)^5$ flavor symmetry is taken to arise from the SM Yukawa couplings. The higher-dimension operators incorporating new physics beyond the SM and breaking the $SU(3)^5$ flavor symmetry, are
suppressed by these factors.

 In our model, with only an $SU(3)^3$ global flavor symmetry and two bi-fundamental spurions to implement the breaking,
we have the special relation $Y_d \propto Y_u \,Y_e^\dagger$ among the Yukawa matrices for the quarks and charged leptons.
In this section, we examine the resultant EFT constructed from SM fields, focusing on the higher-dimension operators that induce flavor-changing neutral currents. As in MFV, we build these operators from only the Yukawa matrices.

We construct them from dimension-three building blocks bilinear in the fermion fields,
containing enough insertions of the Yukawa matrices $Y_e$ and $Y_u$ to respect the
flavor symmetry. For simplicity, we assume here that each building block
by itself respects baryon and lepton number symmetry. One could relax this assumption, and
include higher order operators that contain lepto-quark bilinears. As we will see, the bounds
we find are already quite restrictive, sufficient to reach our main conclusions. The product $Y_u Y_e^{\dagger}$, when encountered, is replaced by $Y_d$ as dictated by Eq.~(\ref{Eq:relation}) to have a more conservative FCNC constraint. Each fundamental bilinear is then taken to enter with no further suppression by small parameters.

We ignore restrictions coming from the $SU(2)_L\times U(1)_Y$
gauge symmetry because we are now implementing the EFT below the EW breaking scale (the Kaon or B-meson scales, for example). We express these building blocks, without contracting color or spinor indexes. Thus several new higher-order operators having the same flavor structure, but different QCD or Lorentz structure, are treated on the same footing.
 We do not list here bilinears such as $\bar{Q}_{L} Q_L$ or $\bar{Q}_L Y_d d_R$
 since it is always possible, at the low energies considered here,
 to write the theory in a basis in which the kinetic and mass terms are diagonal in flavor space.

 The list of possible bilinears includes those that appear in MFV, such as quark bilinears constructed using only $Y_u$ and $Y_d$. It also includes bilinears that do not appear in MFV: quark bilinears  involving factors of $Y_e$ and lepton bilinears involving $Y_u$ and $Y_d$.  In Table~\ref{tab:operators}, we list these bilinears, using only a minimal number of Yukawa matrices of each type. Columns $1$ and $2$ contain quark bilinears appearing in MFV. Those in column $2$, containing  more factors of $Y_d$ (or $Y_e$) in each bilinear, are suppressed relative to those in column $1$ in our single-Higgs-doublet model. With two Higgs doublets and large tan$\beta$, some of them can be enhanced to the level of the column-$1$ bilinears. In column $3$, we list bilinears that do not appear in MFV.

\begin{table}
 \renewcommand{\arraystretch}{1.2}
  \centering
 \scriptsize
  \tabcolsep=0.0cm
\begin{tabular}{|c|c|c|}
\hline\hline
  MFVu & MFVd & CFB\cr
\hline
$O_{1u}=\bar{Q}_L\,Y_uY_u^{\dagger}Q_L$   &
 $O_{1d}=\bar{Q}_L\,Y_dY_d^{\dagger}Q_L$ &
  $O_{1e}=\bar{Q}_L\,Y_dY_eY_e^{\dagger}Y_d^{\dagger}Q_L$\cr
$O_{2u}=\bar{Q}_L\,Y_uY_u^{\dagger}Y_uu_R$ &
$O_{2d}=\bar{Q}_L\,Y_dY_d^{\dagger}Y_uu_R$ &
$O_{2e}=\bar{Q}_L\,Y_dY_eu_R$\cr
$O_{3u}=\bar{Q}_L\,Y_uY_u^{\dagger}Y_dd_R$ &
$O_{3d}=\bar{Q}_L\,Y_dY_d^{\dagger}Y_dd_R$ &
$O_{3e}=\bar{Q}_L\,Y_d\,Y_e Y_e^{\dagger}d_R$\cr
$O_{4u}=\bar{u}_R\,Y_u^{\dagger}Y_u {u}_R$ &
$O_{4d}=\bar{u}_R\,Y_u^{\dagger}Y_dY_d^{\dagger}Y_u {u}_R$ &
$O_{4e}=\bar{u}_R\,Y_e^{\dagger}Y_e {u}_R$ \cr
& $O_{5d}=\bar{d}_R\,Y_d^{\dagger}Y_d {d}_R$ &
$O_{5e}=\bar{d}_R\,Y_eY_e^{\dagger} {d}_R$\cr
& $O_{6d}=\bar{u}_R\,Y_u^{\dagger}Y_d{d}_R$ &
$O_{6e}=\bar{u}_R\,Y_e^{\dagger}{d}_R$\cr
& $O_{1e} = \bar{L}_L Y_e Y_e^\dagger L_L$ & $O_{1\ell} =  \bar{L}_L Y_d^\dagger Y_d L_L$ \cr
& $O_{2e} = \bar{L}_L Y_e Y_e^\dagger Y_e e_R$ & $O_{2\ell} = \bar{L}_L Y_d^\dagger Y_u e_R$ \cr
& $O_{3e} = \bar{e}_R Y_e^\dagger Y_e e_R$ & $O_{3\ell} = \bar{e}_R Y_u^\dagger Y_u e_R$ \cr
\hline\hline
\end{tabular}
\caption{The building blocks (bilinear in SM quarks and charged leptons) contributing to higher order operators.}
\label{tab:operators}
\end{table}

Not all the building blocks induce FCNC processes. In MFV with a single Higgs doublet, only bilinears quadratic in $Y_u$
are important for FCNC processes (bilinears $O_{1u}$ and $O_{3u}$).
It turns out that $O_{1u}$ for example contributes to all the main FCNC transitions mediated by operators of dimension 6
---such as $\Delta m_K$, $\epsilon_K$, $\Delta M_{B_s}$, $A_{CP}(B_d\rightarrow J/\psi \,K_S)$
and $\Delta M_{B_d}$---
while $O_{3u}$ is important in processes dominated by dimension-5 operators (such as $b\rightarrow s\,\gamma$ and
$B_s\rightarrow \mu^+\mu^-$).
The bilinears $O_{2u}$ and $O_{4u}$  do not yield FCNC transitions.

All of the bilinears of columns $1$ and $2$ have been analyzed previously, coming to the 
conclusion when they are employed to construct dimension-$5$ and dimension-$6$ FCNC
 operators, the scale can be as low as the TeV scale.

\subsection{CFB Quark Bilinears}

We next consider the bilinears of column 3, which do not arise in MFV. The six quark-bilinears involve one or more factors of $Y_e$, each of which introduces new off-diagonal structures relative to MFV. One of these operators is harmless since it contributes only to charged currents:
 $O_{6e}$ might look unsuppressed, but in practice
it is subdominant in comparison to the SM tree-level contributions that it competes against.

Operators such as $O_{1e}$, $O_{3e}$, $O_{4e}$ and $O_{5e}$
each contain insertions of $Y_eY_e^{\dagger} $, for which
\beqs
|Y_eY_e^{\dagger}| &=&\frac{1}{\eta^2}
\left(\begin{array}{ccc}
3.4& 14.9 & 11.2\cr
14.9 & 66.2 & 49.6\cr
11.2 & 49.6 & 37.1
\end{array}\right)\,,
\label{Eq:ye-square}
\eeqs
where $\eta^2 \simeq 10^{6}$. They are therefore relatively suppressed.

 As an example, consider the $K^0-\bar{K}^0$ mixing.
 The operator to be considered is $\frac{1}{\Lambda^2}O_{5e}O_{5e}$, with $\Lambda$ a  parameter 
 characterising the scale of new physics, while operators built from $O_{1e}$ and $O_{3e}$
 have additional suppression due to $Y_d$. The scale $\Lambda$ is expected to be given by the flavor-breaking scale $F$ of Section~\ref{sec:framework}, up to model-dependent factors.

 Assuming the same value for the matrix element as the SM operator, and the formalism in~\cite{Buras}, we find
 \beqs
\frac{\Delta M_K({\rm new})}{\Delta M_K({\rm SM})}
&\simeq&\left(\frac{44 \,{\rm GeV}}{\Lambda}\right)^2\,\cdot \left[\frac{(Y_eY_e^{\dagger})_{21}}{10^{-5}} \right]^2\,.
\eeqs
Because $(Y_eY_e^{\dagger})_{21}\sim 10^{-5}$, this observable receives small corrections, negligible for
$\Lambda > 1$ TeV as generically expected already from the MFV operators.

The bilinear $O_{2e}$ contributes for example to $D^0-\bar{D}^0$ mixing through $O_{2e}O_{2e}$. Each bilinear involves  the matrix
\beqs
|Y_dY_e| &=&\frac{1}{\eta}
\left(\begin{array}{ccc}
3\times 10^{-5}& 2\times10^{-8}& 2\times 10^{-12}\cr
0.003 & 3\times 10^{-5} & 5\times 10^{-9}\cr
0.1 & 0.003 & 0.0003
\end{array}\right)\,,
\eeqs
where the $21$ entry is very small. Again this process is adequately suppressed with a scale on the order of 1 TeV.

We conclude that the new operators arising in CFB are such that all quark FCNC 
processes are adequately suppressed by flavor scales as low as 1 TeV --- just as in MFV.

\subsection{CFB Charged-Lepton Bilinears}
Of the three lepton bilinears of column-$3$, $O_{1\ell}$ is suppressed by at least two powers of the bottom Yukawa coupling and is harmless. The second operator $O_{2\ell}$, though, because of large off-diagonal entries in $Y_d^{\dagger}Y_u$, does introduce a stringent constraint on the flavor scale.
Diagonalizing $Y_e$ via $Y_e=L^{(e)} \mbox{diag}(Y_e) R^{(e)\dagger}$ and using the relation in Eq.~(\ref{Eq:relation}), we have the flavor-changing matrix among the lepton mass eigenstates as
\beqs
&&\hspace{-0.8cm}\eta\, |\mbox{diag}(Y_e)\,R^{(e)\dagger}\, \mbox{diag}(Y_u)^2\,R^{(e)}|  \nonumber \\
&=&\eta\, \left(\begin{matrix}
  2.2\times 10^{-6} & 2.3\times 10^{-7} & 2.1\times 10^{-9} \\
 1.3\times 10^{-5} &  1.4\times 10^{-6}  & 1.3\times 10^{-8} \\
9.9\times 10^{-6} & 1.0\times 10^{-6} &  9.7\times 10^{-9}
 \end{matrix} \right)  \;.
 \label{eq:operator-2}
\eeqs
Dressing this operator with a Lorentz structure, we have $1.3\times 10^{-2}\, v_{W}\,\bar{\mu}_L \sigma^{\mu\nu} e_R F_{\mu\nu}/(\sqrt{2}\,\Lambda^2)$ or $9.9 \times 10^{-3} \,v_{W}\,\times\bar{\tau}_L \sigma^{\mu\nu} e_R F_{\mu\nu}/(\sqrt{2}\,\Lambda^2)$ or $1.0 \times 10^{-3}\,v_{W}\, \times\bar{\tau}_L \sigma^{\mu\nu} \mu_R F_{\mu\nu}/(\sqrt{2}\,\Lambda^2)$. The currently most stringent constraint comes from the rare process $\mu \rightarrow e \gamma$, requiring  $\Lambda \gtrsim 10^4$~TeV~\cite{LFV},~\footnote{Our constraint is fairly conservative in a sense that an operator constructed from the spurions, $Y_u$ and $Y_e$, will not have this additional factor of $\eta$ in  Eq.~(\ref{eq:operator-2}).} far beyond the energy reached by the Large Hadron Collider. 
The operator $O_{2\ell}$ also leads to a simple relation between the three radiative decays
$\mu\rightarrow e\gamma$, $\tau\rightarrow \mu\gamma$ and $\tau\rightarrow e \gamma$
that might become of relevance for future experimental searches~\cite{Abrams:2012er}.

Another constraint arises from the bilinear $O_{3 \ell}$. After rotating to the mass eigenstate basis, we have the flavor-changing matrix
\beqs
&& \hspace{-0.8cm} |R^{(e)\dagger }  \, \mbox{diag}(Y_u)^2 \, R^{(e)} |  \nonumber \\
&=& \left(\begin{matrix}
  0.99 & 0.10 & 1.0\times 10^{-3} \\
0.10 &  0.01  & 1.0\times 10^{-4} \\
1.0\times 10^{-3} & 1.0\times 10^{-4} &  1.0\times 10^{-6}
 \end{matrix} \right)  \,.
\eeqs
The relatively large $12$ and $21$ entries are again a consequence of large off-diagonal entries in $Y_e$. They lead to a bound on the flavor scale via the process  $\mu^{-} \rightarrow e^-e^+e^-$. Employing the dimension-six operator, $0.1\,\bar{\mu}_R \gamma_\mu e_R \, \bar{e}\gamma^\mu e/\Lambda^2$, the constraint  is $\Lambda \gtrsim 320$~TeV~\cite{LFV}.

Finally we note that the dimension-$6$ operators formed by composing the bilinears of Table~\ref{tab:operators} can also 
be used at the quantum level, where they can lead to loop corrections to Eq.~(\ref{Eq:relation}).
Composing the bilinear $O_{1u}$ with itself, for example, and employing this operator twice, produces a 
two-loop correction to the down-quark mass operator $ H \bar{Q}_L  Y_d  d_R$, and therefore to Eq.~(\ref{Eq:relation}). 
It is of order $M^4 / (4\pi \Lambda)^4$ 
times factors of $Y_u$ and $Y_u^{\dagger}$, where $M$ is the cutoff (the scale of new physics). 
As long as $M\lesssim\Lambda$, this will
amount to a small correction to Eq.~(\ref{Eq:relation}).

\section{Discussion and conclusions}
\label{sec:conclusions}

We have searched for predictive flavor models based on subgroups of 
$SU(3)^5$, the full flavor group of the quarks and charged leptons in 
the standard model. Restricting to products of $SU(3)$'s,
we were led to a model based on $SU(3)^3$, with its breaking due to two, 
rather than three, independent $3\times 3$ spurion fields (Yukawa 
matrices). Thus one of the three
Yukawa matrices $Y_u$, $Y_d$ and $Y_e$ can be expressed in terms of the 
other two. Taking $Y_u$ and $Y_e$ to be
the basic spurion fields, we chose the symmetry assignments of the 
fermions and the spurions in such a way as to realize the relation 
$Y_d\propto Y_uY_e^{\dagger}$.

This leads to an interesting relation between the 
charged-lepton mass ratios and quark mass ratios and mixing angles, 
shown in Eqs.~(\ref{Eq:approx-formula-1}) and (\ref{Eq:approx-formula-2}). It 
also leads to an unusual, strongly off-diagonal structure of the Yukawa 
matrix $Y_e$, indicating that large values of the PMNS angles are 
natural. While we have not yet incorporated neutrino mass into our 
model, it would require a fine-tuned relation between $Y_e$ and the 
neutrino mass matrix to avoid the presence of large PMNS angles. We have 
provided approximate analytic expressions for the charged lepton mass 
ratios in Eqs.~(\ref{Eq:approx-formula-1}) and (\ref{Eq:approx-formula-2}), 
noting their dependence on quark masses ratios, and commenting on the 
adjustments that would be required to bring $m_{\mu}/m_{\tau}$ into 
closer agreement with experiment. The strongly off-diagonal structure of $Y_e$ also plays a role in making the flavor scale $\Lambda$ much larger than in MFV.

These results emerge from an EFT incorporating the standard model and the higher-dimension
operators that correct it. The results follow from the postulated flavor-symmetry 
structure $Y_d\propto Y_uY_e^{\dagger}$.  This relation is stable in the 
presence of SM radiative corrections since these interactions are 
suppressed by the small couplings of the SM. It is stable even in the 
framework of the UV completion of the EFT, up to the flavor breaking 
scale $F$, providing that these interactions, too, are weakly coupled at 
this scale. The important underlying issue of the 
stability of the electroweak scale itself has to be addressed
by additional ingredients that we have not discussed here.

Our study is motivated by minimal flavor violation, which also is based
on symmetry arguments alone without specifying the UV dynamics involved
in flavor generation. Our approach, though, is more restrictive,
postulating a smaller broken symmetry in order to obtain predictive
relations among the quark and charged-lepton masses and mixing angles.
This goal is partly met, with the conclusion that leptonic FCNC
processes constrain the general flavor-symmetry-breaking scale to be
above $10^4$ TeV, well above the energies accessible at the LHC.
This appears to confirm a general lesson: more restrictive generalizations of MFV,
providing some predictivity without reference to a UV completion, require such large scales.
We also found an unexpected result:  large mixing angles in the leptonic sector
arise naturally, again without explicit reference to the UV completion.
In our particular model, FCNC effects are very small in the quark sector
but potentially observable in the next generation of charged-lepton
flavor-violating experiments, for example the Fermilab Mu2e
experiment~\cite{Abrams:2012er}.

\vspace{1.0cm}
\begin{acknowledgments}
We thank Aneesh Manohar, Robert Shrock and Witold Skiba for helpful discussion.
The work of TA is supported by the U. S. Department of Energy under the contract DE-FG02-92ER-40704. The work of YB is supported by the U. S. Department of Energy under the contract DE-FG02-95ER-40896. The work of MP is supported in part by WIMCS and by the STFC grant ST/L000369/1.  We thank the Aspen Center for Physics, under NSF Grant No. PHY-1066293, where some of this work was done. YB also would like to acknowledge the Mainz Institute for Theoretical Physics (MITP) for its hospitality during the completion of this work.
\end{acknowledgments}

\appendix

\section{Notation and numerical input}
\label{sec:notation}

We collect in this Appendix all the relevant expressions
that fix the notation and briefly summarise numerical values for the experimental 
quantities of relevance to the paper, taken from~\cite{Agashe:2014kda}.

We write the dimension-3 mass terms  for the charged fermions
of the SM in the form $-\bar{\psi}^{(i)}_L M^{(i)}\psi_R^{(i)}$,
with $i=e,d,u$.
With these conventions, the diagonalization of the mass matrices can be done by introducing
three unitary matrices  $L^{(i)}$ acting the left-handed fields and three unitary matrices $R^{(i)}$
acting on the right-handed fields.
We use the conventions according to which
\beqs
{\rm diag}\,M^{(i)}&=&L^{(i)\,\dagger} M^{(i)}R^{(i)}\,,
\eeqs
which imply that in the mass basis the charged-current weak interactions contain the CKM
mixing matrix
\beqs
V_{\rm CKM}&=&L^{(u)\,\dagger}L^{(d)}\,.
\eeqs

These definitions imply for example that if we choose to write the Lagrangian in a basis in which
weak interaction are diagonal (flavour basis), and in which the down-type quarks are
diagonal, then the up-type quark flavour eigenstates $|u\rangle$ are related to the
mass eigenstates $|\hat{u}\rangle$ as
\beqs
\label{Eq:CKM}
|\hat{u}\rangle &=& V_{\rm CKM} |u\rangle\,.
\eeqs

The conventional way to parameterise a unitary matrix $V_{\rm CKM}$
requires to introduce three angles $\theta_{ij}$ and a CP-violating phase $\delta$.
In terms of these, we write
\begin{widetext}
\beqs
V_{\rm CKM}&=&
\left( \begin{array}{ccc}
c_{12}c_{13}  &  s_{12}c_{13}  &  s_{13} e^{-i\delta}  \cr
-s_{12}c_{23}-c_{12}s_{23}s_{13}e^{i\delta}  & c_{12}c_{23}-s_{12}s_{23}s_{13}e^{i\delta} & s_{23} c_{13} \cr
s_{12}s_{23}-c_{12}c_{23}s_{13}e^{i\delta}  & -c_{12}s_{23}-s_{12}c_{23}s_{13}e^{i\delta}  & c_{23}c_{13}
\end{array} \right)\,,
\eeqs
\end{widetext}
where $s_{ij}\equiv \sin\theta_{ij}$ and $c_{ij}\equiv\cos\theta_{ij}$. The experimental values are
\beqs
\sin\theta_{12}&=&0.2243\pm0.0016\,, \nonumber\\
\nonumber
\sin\theta_{23}&=&0.0413\pm0.0015\,,\\
\nonumber
\sin\theta_{13}&=&0.0037\pm0.0005\,, \\
\nonumber
\sin{2\beta} &=&0.679\pm 0.019\,.
\eeqs
The angle $\beta \equiv \mbox{arg}[- V_{cd} V^*_{cb}/(V_{td} V^*_{tb})]$ in the CKM  unitarity triangle is related to the phase $\delta$.

For the CKM matrix, one could also use the Wolfenstein parametrization,
\beqs
V_{\rm CKM} = \left[
\begin{matrix}
1 - \lambda^2/2  & \lambda   & A \, \lambda^3 (\rho - i \,\eta)    \\
- \lambda  & 1  - \lambda^2/2   & A \, \lambda^2 \\
A \,\lambda^3 ( 1- \rho - i\,\eta)   & - A\, \lambda^2   & 1
\end{matrix}
\right] \,.
\label{Eq:Wolfenstein}
\eeqs

The experimental values are 
\beqs
\nonumber
\lambda &=& 0.22535\pm 0.00065\,,\\
\nonumber
A &=& 0.811^{+0.022}_{-0.012}\,,\\
\nonumber
\bar{\rho} &\approx& \rho \,=\, 0.131^{+0.026}_{-0.013}\,,\\
\nonumber
\bar{\eta}&\approx& \eta \,=\, 0.345^{+0.013}_{-0.014}\,.
\eeqs

Finally, for the masses of the quarks we adopt the convention of considering all of them as defined at the
common scale $M_Z$. We have
\beqs
\nonumber
m_t&=& 176\pm 5\,\, {\rm GeV}\,,\\
\nonumber
 m_b&=& 2.95\pm 0.15\,\, {\rm GeV}\,,\\
\nonumber
m_c&=& 0.65\pm 0.12\,\, {\rm GeV}\,,\\
\nonumber
m_s&=& 0.062\pm 0.015\,\, {\rm GeV}\,,\\
\nonumber
m_u&=& 0.0017\pm 0.0005\,\, {\rm GeV}\,,\\
\nonumber
m_d&=& 0.0032\pm 0.0009\,\, {\rm GeV}\,.
\eeqs
The masses of the charged leptons are affected by smaller errors and smaller
running effects, both of  which we neglect,
and are given by
\beqs
\nonumber
m_{\tau}&=&1.78\,\,{\rm GeV}\,,\\
\nonumber
m_{\mu}&=&0.106\,\,{\rm GeV}\,,\\
\nonumber
m_e&=&0.511\,\times\,10^{-3}\,\,{\rm GeV}\,.
\eeqs


\end{document}